\documentclass{article}

\usepackage[preprint]{neurips_2023}

\usepackage[utf8]{inputenc} 
\usepackage[T1]{fontenc}    
\usepackage{hyperref}       
\usepackage{url}            
\usepackage{booktabs}       
\usepackage{amsfonts}       
\usepackage{nicefrac}       
\usepackage{microtype}      
\usepackage{xcolor}         
\usepackage{amsfonts}
\usepackage{graphicx}
\usepackage{makecell}
\usepackage{booktabs} 
\usepackage{subfig}
\usepackage{amsmath}
\usepackage{accents}
\usepackage{statex}
\usepackage[normalem]{ulem}
\usepackage{enumitem}
\usepackage{multirow}
\usepackage[nomargin,inline,marginclue,draft]{fixme}
\usepackage{balance}
\usepackage{changepage}
\usepackage{bm}
\usepackage{hyperref}
\usepackage{setspace}
\usepackage{mathrsfs}
\usepackage{ulem}
\usepackage{verbatim}
\usepackage{diagbox}
\usepackage{pdftexcmds}
\usepackage{catchfile}
\usepackage{ifluatex}
\usepackage{ifplatform}
\usepackage{threeparttable}
\usepackage{comment}
\usepackage{algorithm}
\usepackage{algorithmicx}
\usepackage{algpseudocode}
\usepackage{wrapfig}
\usepackage{color}
\usepackage{colortbl}

\title{Identify Critical Nodes in Complex Network with Large Language Models}

\author{%
  Jinzhu~Mao$^{1}$,
  Dongyun~Zou$^{1}$, 
  Li~Sheng$^{1}$,
  Siyi~Liu$^{2}$,
  Chen~Gao$^{1}$,
  Yue~Wang$^{1}$,
  Yong~Li$^{1}$\\ 
  {$^{1}$Department of Electronic Engineering, BNRist, Tsinghua University, China}\\
  {$^{2}$École Polytechnique Fédérale de Lausanne, Switzerland} \\
    \small{\texttt{\{maojz22,zoudy22,chengl22\}@mails.tsinghua.edu.cn}}\\
    \small{\texttt{\{ssui.liu1022,chgao96\}@gmail.com}}\\
    \small{\texttt{\{wangyue,liyong07\}@tsinghua.edu.cn}} \\
  \vspace{-5mm}
}

\begin{document}
\maketitle
\begin{abstract}
    Identifying critical nodes in networks is a classical decision-making task, and many methods struggle to strike a balance between adaptability and utility. Therefore, we propose an approach that empowers Evolutionary Algorithm (EA) with Large Language Models (LLMs), to generate a function called "score\_nodes" which can further be used to identify crucial nodes based on their assigned scores. 
    Our model consists of three main components: Manual Initialization, Population Management, and LLMs-based Evolution. It evolves from initial populations with a set of designed node scoring functions created manually. LLMs leverage their strong contextual understanding and rich programming skills to perform crossover and mutation operations on the individuals, generating excellent new functions. These functions are then categorized, ranked, and eliminated to ensure the stable development of the populations while preserving diversity.
    Extensive experiments demonstrate the excellent performance of our method, showcasing its strong generalization ability compared to other state-of-the-art algorithms. It can consistently and orderly generate diverse and efficient node scoring functions. All source codes and models that can reproduce all results in this work are publicly available at this link: \url{https://anonymous.4open.science/r/LLM4CN-6520}
\end{abstract}

\section{Introduction}

Various networks widely exist in the real world, such as social networks, biological networks, etc~\cite{west2001introduction}.
For the network, identifying an optimal set of critical nodes for which the activation/removal has the maximized impact is a basic research problem with broad applications in epidemic control, influence maximization, etc~\cite{lalou2018critical,arulselvan2009detecting}.
From the perspective of combinational optimization, this problem could be considered as a kind of optimal search task, in which the computational complexity increases exponentially along with the size of the network \cite{taniguchi2006critical}.
Existing approaches to this problem can be categorized as follows: from heuristic methods to optimization-based methods, and finally, to learning-based methods.

The first kind of work is the basic methods based on heuristics, such as using the Degree Centrality \cite{zhang2017degree}, to judge whether the node is critical or not, which is simple yet very effective and efficient, showing superior performance in many kinds of networks ~\cite{arulselvan2009detecting,mao2023detecting}.
The second kind of work further tries to re-formulate the problem with approximated optimization problems.
Some other algorithms like CoreHD \cite{CoreHD:zdeborova2016fast} and WeakNeighbors \cite{WeakNeighbors:schmidt2019minimal} utilize \textit{k}-core decomposition, in which the \textit{k}-core of a network is defined as the maximal subgraph with every node having a degree of at least $k$. These methods progressively eliminate nodes with the highest degrees within their respective \textit{k}-cores to identify key nodes. 
Furthermore, Min-Sum \cite{minsum:braunstein2016network} focuses on creating an acyclic network by removing nodes that form loops, followed by utilizing greedy tree-breaking algorithms to decompose the remaining forest into smaller disconnected components. 
Another work, GND \cite{GND:ren2019generalized}, adopts a spectral approach to iteratively partition the network into two components. 
These methods have the advantage of fast computation, but they may lead to sub-optimal solutions, or even lack a suitable solution altogether.
Recently, with the advancement of deep learning, some works~\cite{NIRM:zhang2022dismantling, GDM:grassia2021machine} approach the problem by using neural networks to learn the node importance or ranking nodes adaptively. That is, they utilize neural networks to assign scores to network nodes, determining their importance based on these scores.

Despite these advances, there still exists a key challenge of \textit{adaptability-utility dilemma}, which largely affects the real application values of these methods.
Specifically, the heuristic or optimization methods provide the same solution for all kinds of networks, but however, ignoring the different characteristics, which leads to poor performance in those unpopular networks.
The learning-based methods try to search the critical nodes adaptively, but are faced with high search costs in such a huge search space, which in turn, leads to poor utility. 

Recently, Large Language Models (LLMs) have shown a human-like ability in many tasks, from basic natural language processing, to code generation, and some high-level logical reasoning and decision-making tasks~\cite{zhao2023survey,chang2023survey,gao2023large}.
Inspired by these advances, we propose a novel solution that adopts LLMs to generate the solution and design an evolutional learning framework to fully unleash LLMs.
Specifically, we utilize LLMs-empowered evolutionary algorithms to address the critical node problem, with LLMs performing crossover and mutation operations, continuously iterating to generate new solutions.

The contributions of this work can be summarized as follows:
\begin{itemize}[leftmargin=*]
    \item We approach the fundamental problem of critical node detection from a brand new perspective of LLMs, for which the original problem is transformed into a code-generation task.
    \item We propose a general framework integrating LLMs and evolutional learning. We propose a generation-updation mechanism for LLMs and the crossover-mutation method for evolutional learning.
    \item We conduct extensive experiments on both real-world and synthetic networks, and the results show the superior performance of our proposed method. The searched solutions can further inspire researchers and practitioners in this area.
\end{itemize}

\section{Problem Statement}\label{sec::profdef}

In this work, our objective is to learn a function $f$ that can identify critical nodes within a network. 

Given a network $\mathcal{G}=(\mathcal{V}, \mathcal{E})$, where $\mathcal{V}$ represents the set of nodes $\{v_1, v_2, ... , v_V\}$ and $\mathcal{E}$ represents the set of edges $\{e_1, e_2, ... , e_E\}$, we aim to identify a set of critical nodes $\mathcal{V}_c$ = $\{v_{c_1}, v_{c_1}, ... , v_{c_L}\}$, such that when these nodes are removed or attacked, the network experiences significant disruptions. 
For the sake of facilitating function reading, we represent the network $\mathcal{G}$ as an adjacency matrix $ \mathbf{A}=\{0, 1\}^{V\times V}$, where $ \mathbf{A}_{jk}=1$ indicates that there is an edge between node $v_j$ and node $v_k$; otherwise, $\mathbf{A}_{jk}=0$.

Many existing methods rely on a crucial parameter, which is the number of target nodes $N$, directly impacting the algorithm's performance.
In order to ensure the efficiency and generalization of the function learned by our approach, our objective function $f$ does not include the variable $N$. 
Instead, it will take the network's adjacency matrix $\mathbf{A}$ as input and produce scores for each node $\mathcal{S}= \{s_{v_1}, s_{v_2}, ... , s_{v_V}\}$ as output. In other words, $\mathcal{S} = f(\mathbf{A})$, which enables us to rank nodes based on their scores and obtain any desired quantity of critical nodes without specifying the number in advance. 

This node scoring function is obtained through our proposed evolutionary algorithm empowered by large model models. 
Based on the given initial functions $\mathcal{F}_0 = \{f_{0_1}, f_{0_2}, ... , f_{0_M}\}$, our model generates some new functions during the training process through crossover and mutation.
To assess the fitness of these functions, we introduce an evaluation algorithm $\operatorname{E}$, which determines whether a function's scores for nodes are reasonable, enabling us to update and eliminate functions accordingly. 
Therefore, the evaluation algorithm $e$ explicitly defines the significance of "critical nodes" within the network.
In the end, many excellent node scoring functions $ \mathcal{F}_T = \{f_{T_1}, f_{T_2}, ... , f_{T_N}\}$ are retained within the population, and we select the one with the highest fitness as the model's output.

\section{Methodology}\label{sec::method}

\begin{figure*}[t]
  \vspace{-1em} 
  \centering
  \includegraphics[width=1\textwidth]{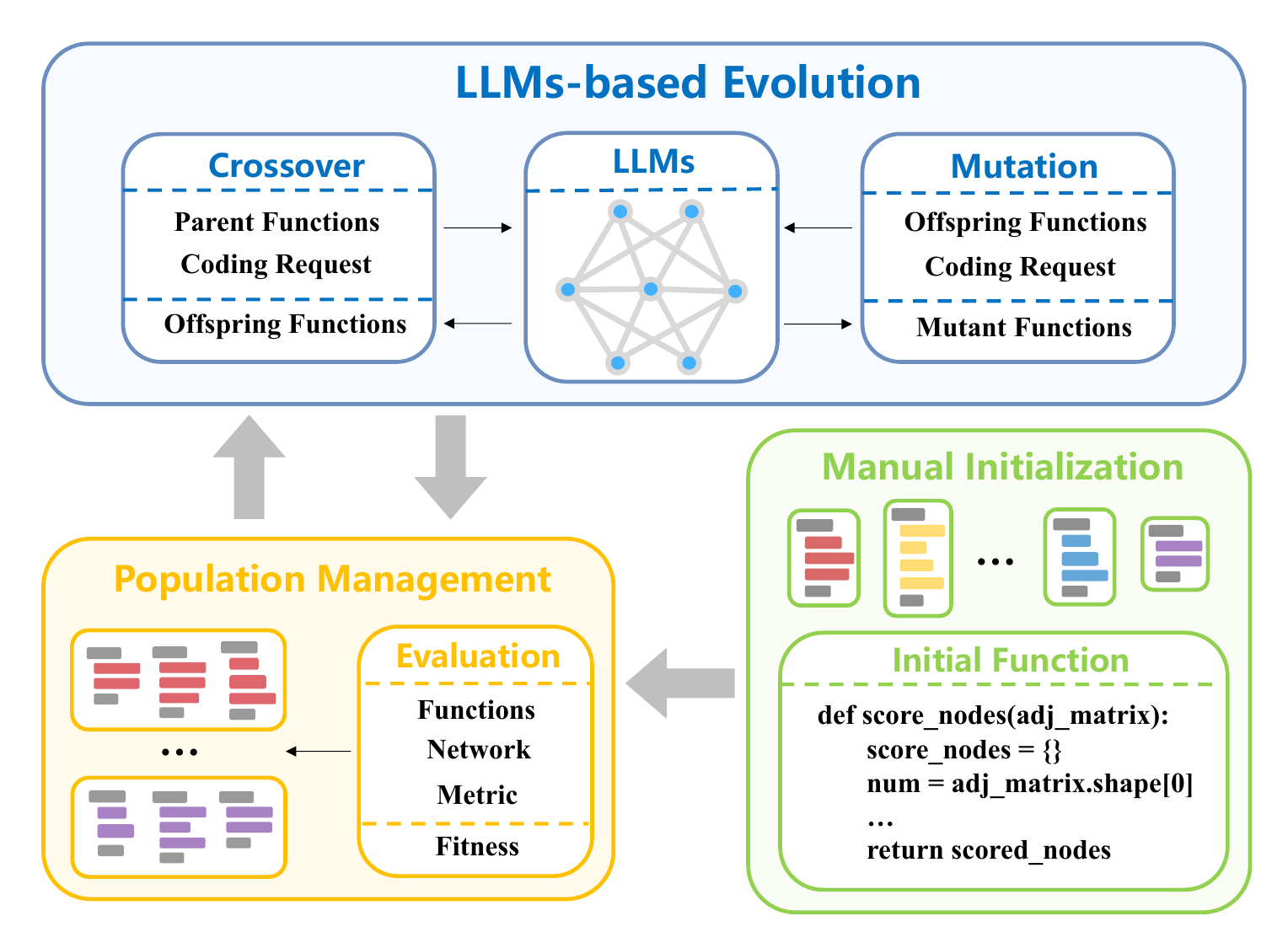}
  \vspace{-2em} 
  \caption{Illustration of our proposed model.} 
  \label{fig::framework}
  \vspace{-1em} 
\end{figure*}

In order to discover a brand new excellent node scoring function, our model is based on evolutionary algorithms, combined with the powerful contextual understanding and programming capabilities of LLMs. 
Through generation, evolution, evaluation, and iteration, we obtain a novel high-performance function. 
It primarily consists of three components: Manual Initialization, LLMs-based Evolution, and Population Management. 
Figure \ref{fig::framework} and Algorithm \ref{alg:framework} illustrate the framework of our model.

\subsection{Manual Initialization}\label{sec::Init}
Evolutionary algorithms is a class of population-based stochastic search strategies. 
The population initialization provides potential individuals for the entire search process.
These initial individuals continually iterate and improve during the training process, getting closer to the optimal solution. 
A well-designed initialization can assist evolutionary algorithms in finding the optimal solution, particularly when dealing with large-scale optimization problems using finite-sized populations. 
Therefore, we categorize the design approaches for node scoring functions into two classes as follows.
\begin{itemize}[leftmargin=*]
\item \textbf{Network Topology-based Function:} 
Network topology encompasses the connectivity relationships among nodes, and critical nodes often exhibit distinct connectivity patterns compared to other nodes. 
In this section, we assign corresponding scores to nodes based on their centrality metrics, such as the number of N-hop neighbors of a node, the number of shortest paths in which a node lies, and the reciprocal of the sum of shortest paths from a node to all other nodes.
Because of the availability of comprehensive library functions, some centrality metrics can now be directly utilized. 
To achieve a balance between the scalability and readability of the node scoring function, we choose not to rely solely on concise library functions for node scoring; instead, we have designed some functions that include detailed computation processes.

\item \textbf{Algorithmic Method-based Function:} 
Many classical algorithms analyze network characteristics through mathematical or statistical methods, and they identify critical nodes to dismantle the network based on the results of analysis.
These methods often incorporate greedy algorithms, the efficiency of which is closely related to the network size and the number of selected nodes. 
In this work, we retain the core ideas of these algorithms and manually rewrite them into an efficient node scoring function as our initial approach. 
It is worth noting that we did not consider learning-based methods because they typically require significant computational resources, which contradicts our goal of having a concise, clear, and fast node scoring function.
\end{itemize}
Based on the two approaches described above, we designed $M$ initial functions $\mathcal{F}_0 = \{f_{0_1}, f_{0_2}, ... , f_{0_M}\}$, each with a different emphasis in scoring nodes, all of which can serve as independent populations $ \mathcal{P} = \{P_1, P_2,..., P_M\}$ to start evolution. The input and output of the node scoring function have been defined in Section \ref{sec::profdef} as $f(\mathbf{A}) = \{s_{v_1}, s_{v_2}, ..., s_{v_V}\}$.

\subsection{Population Management}\label{sec::Population}
To maintain diversity within the function population, avoid falling into local optima, and simultaneously improve the performance of functions, we implement a sophisticated management strategy for the function population.

\begin{algorithm}[t]
\caption{Algorithm of LLMs+EA}
\label{alg:framework}
\begin{algorithmic}[1] 
\State \textbf{Input:} Initial function set $\mathcal{F}_0$; Evolution epochs $T$; Mutation probability $\theta$; Adjacency matrix $\mathbf{A}$; Prompt for Crossover $prompt_{cross}$, Prompt for mutation $prompt_{mutate}$; Similarity threshold $\tau$
\State \textbf{Output:} Best function $f^*$.
\State $\mathcal{F}_0$ as the independent population 
\For{$epoch = 1$ to $T$} 
    \ForAll{$f$ in $\mathcal{F}_t$}
        \State Evaluate fitness score $c_f$ of $f$ on $\mathbf{A}$;
        \State Classify $f$ to the most similar population or create a new one based on $\tau$;
    \EndFor        
    \State Sample functions $\mathcal{F}_{intra}$ and $\mathcal{F}_{inter}$ as $\mathcal{F}_{parent}$ from all functions;
    \State Generate offspring functions $\mathcal{F}_{off}$ using LLMs given $\mathcal{F}_{parent}$ and $prompt_{cross}$;
    \State Sample from $\mathcal{F}_{off}$ with probability $\theta$ and input with $prompt_{mutate}$ into LLM to get $\mathcal{F}_{mutate}$;
    \State Combine Residual $\mathcal{F}_{off}$ and $\mathcal{F}_{mutate}$ as $\mathcal{F}_{t+1}$ for the next evolution round;
\EndFor
\State \Return Function with the highest fitness as $f^*$.
\end{algorithmic}
\end{algorithm}

\subsubsection{Evaluation} 
Given a node scoring function $f \in \mathcal{F}$, our model first conducts a fitness assessment, which is closely related to our specific task --- identifying critical nodes in the network through the learned function. Therefore, we define the node scoring function's fitness evaluation as follows:
\begin{equation}
\operatorname{E}(f) = \sigma\left( \mathbf{A}, \text{SORT}\left(f(\mathbf{A}) \right)[:L] \right) \text {,}
\end{equation}
where \text{SORT} represents a function that sorts nodes in descending order based on their scores, $[:L]$ indicates selecting the top L elements from the sorted list, and $\sigma$ stands for a function that evaluates the network state after removing nodes. 
Through this evaluation function, each executable node scoring function obtains a suitable fitness score $ c_f = \operatorname{E}(f)$, which assists us in better-performing selection and managing the functions population.
\subsubsection{Classification}
While some functions have low fitness, they can introduce diverse potential evolutionary directions to the population. 
Conversely, high-fitness functions already have numerous similar individuals, but their widespread presence often indicates a tendency to get trapped in local optima. 
Considering the above situation, we further classify the node scoring functions $F$ based on their fitness scores.

Inspired by natural language processing tasks, we utilize pre-trained language model \cite{DBLP:journals/corr/abs-1810-04805} to process function $f \in \mathcal{F}$ into embeddings $\mathbf{z}_f$, which contain the syntax structure and semantic information of the code.
Subsequently, we calculate the cosine semantic similarity $\operatorname{S}(\mathbf{z}_f, \mathbf{z}_{P_i})$ between the function and the existing population as follows:
\begin{equation}
\operatorname{S}(z_f, z_{P_i}) = \frac{\mathbf{z}_f \cdot \mathbf{z}_{P_i}}{\|\mathbf{z}_f\| \|\mathbf{z}_{P_i}\|} \text {,}
\end{equation}
\begin{equation}
\mathbf{z}_{P_i} = \frac{1}{|P_i|} \sum_{f \in P_i} \mathbf{z}_f \text {,}
\end{equation}
where $\mathbf{z}_{P_i}$ represents the average of function vectors within population $P_i$. 
At the same time, we establish a similarity threshold $\tau$. Let 
\begin{equation}
i^* = \arg\max_i \operatorname{S}(z_f, z_{P_i})\text {,}
\end{equation}
if $\operatorname{S}(z_f, z_{P_i}) > \tau$, $P_{i^*} = P_{i^*} \cup \{f\}$; 
otherwise, this distinctive function initiates a new population $P_\text{new}$. It is worth noting that each population has a limit on the number of individuals. Once a population reaches its maximum capacity, any newly added function must have a higher fitness score to replace the existing lowest-scoring function, thereby maintaining the population's high performance.

\subsubsection{Selection}
We implemented two designs for selection of parent functions within the population, namely, intra-population selection and inter-population selection.
\begin{itemize}[leftmargin=*]
\item \textbf{Intra-population selection:}
Considering the diversity of functions, our model selects one individual from each population to serve as intra-population parent functions $\mathcal{F}_\text{intra}$ in each selection. At the same time, to maintain elite individuals, the highest-fitness function is also added to $\mathcal{F}_\text{intra}$.

\item \textbf{Inter-population Selection:} 
Taking into account the fitness of functions, two functions are chosen from the same population to serve as inter-population parent functions $\mathcal{F}_\text{inter}$ in each selection, allowing subsequent crossover operations to discover differences between functions of the same population but with different fitness scores.
\end{itemize}
In both selection methods, the fitness scores of each function serve as selection weights, encouraging the selection of more excellent individuals. For convenience, we use $\mathcal{F}_\text{parent}$ to represent  $\mathcal{F}_\text{intra} \cup \mathcal{F}_\text{inter}$.

\begin{wrapfigure}{}{0.5\textwidth}
\vspace{-0.7in}
\begin{center}
\includegraphics[width=0.5\textwidth]{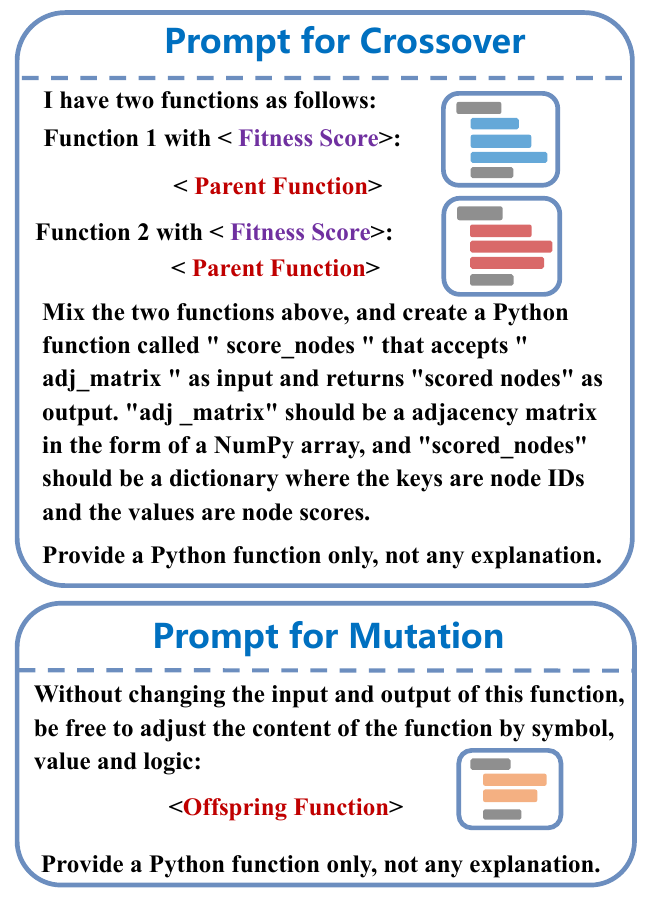}
\end{center}
\vspace{-0.1in}
\caption{Prompts used in the crossover and mutation.}
\label{fig:Prompts}
\vspace{-40pt} 
\end{wrapfigure}

\subsection{LLMs-based Evolution}\label{sec::Evolution}
Given the parent functions, our model begins the evolution process to generate new functions.
The traditional crossover and mutation operators used in programming struggles to make high-quality modifications to functions like human programmers do.
In contrast, LLMs trained on extensive data including code, documentation, tutorials, and more, can comprehend complex programming tasks and perform high-quality code modifications and generation.
Therefore, we use LLMs as the operators to cross and mutate the given functions to generate offspring codes.

\subsubsection{Crossover}\label{sec::Crossover}

The inputs for crossover consist of a set of selected parent functions $\mathcal{F}_\text{parent}$, and the outputs are a set of offspring functions $\mathcal{F}_\text{off}$. 
The prompts provided for LLMs include the functions and their scores, as well as a text-formed description of crossover requirement and additional format-related constraints. We aim for the LLMs to focus on analyzing the distinctions and relationships among functions. Therefore, we have not provided any descriptions related to the network or the task of identifying critical nodes, to avoid causing any interference. The formula for the crossover is shown below:
\begin{equation}
\mathcal{F}_\text{off} = \operatorname{LLM}(\mathcal{F}_\text{parent}, C_\text{parent}, prompt_\text{cross}) {,}
\end{equation}
where $C_\text{parent} = \{c_f, \forall f \in \mathcal{F}_\text{parent} \}$ represent the set of fitness scores of all parent functions.
\subsubsection{Mutation}\label{sec::Mutation}
For the diversity of the population, each offspring function $f \in \mathcal{F}_\text{off}$ has a certain probability $\theta$ of undergoing mutation. The prompts provided as inputs for LLMs include the offspring function and a text-formed description of mutation guiding and format requirements, and the output are a set of mutant functions $\mathcal{F}_\text{mutate}$. Similarly, to keep LLMs focused on the code, our prompts do not contain any descriptions related to the task background. The mutation formula is as follows:
\begin{equation}
\mathcal{F}_\text{mutate} = \operatorname{LLM}(\mathcal{F}_\text{off}, prompt_\text{mutate}){.}
\end{equation}
Figure \ref{fig:Prompts} illustrates the detailed prompts used in the crossover and mutation. The generated functions $\mathcal{F}_\text{off}$ and $\mathcal{F}_\text{mutate}$, after undergoing executable checks, are systematically added to the population following the evaluation and classification described in Section \ref{sec::Population}. This process repeats iteratively, with the entire function population gradually growing until ideal function emerges.

\section{Experiments and Results}\label{sec::experiments}

In this section, we aim to comprehensively evaluate the proposed model by providing detailed responses to the following research questions (RQs):
\begin{itemize}[leftmargin=*]
    \item RQ1: Can our proposed method outperform existing baselines in terms of performance?
    \item RQ2: Which functions does our model discover, and are they identical?
    \item RQ3: Through what processes does our model derive these functions?
    \item RQ4: How effective is each component within our proposed model?
\end{itemize}

\subsection{Experimental Settings}

\subsubsection{Experimental environment}


We compared the performance of our best algorithm obtained from LLMs+EA with numerous traditional methods on both real-world and synthetic networks, including both machine learning-based methods and algorithmic approaches. We contrasted the disruptive impact on the networks during the removal of 20\% of nodes under the ANC metric. The entire model is implemented in Python and runs on Intel Xeon Platinum 8358. We use the hyperparameters in Table \ref{tab:parameters of EA} for LLMs+EA runs, and each dataset requires approximately two dollars with gpt-3.5-turbo for three hours.

\begin{table}[h]
    \centering
    \caption{Hyperparameters of LLMs+EA}
    \label{tab:parameters of EA}
    \begin{tabular}{l|c}
    \toprule
    Hyperparameter& Value \\
    \midrule
    LLM (crossover and mutation)& gpt-3.5-turbo-0613\\
    LLM temperature (crossover and mutation)& 1/1.5 \\
    Number of initial populations&10\\
    Maximum population size&10\\
    Maximum number of epoch&100\\
    Mutation rate& 0.3\\
    Code2Vec model& bert-base-uncased\\
    Similarity threshold& 0.93\\
    \bottomrule
    \end{tabular}
    \vspace{-1em} 
\end{table}

\subsubsection{Metrics} \label{metrics}
Given a network $\mathcal{G=(V, E)}$, with its node set as $\mathcal{V}$ and edge set as $\mathcal{E}$, and a connectivity measure as $\mathcal{\sigma}$, the accumulated normalized connectivity (ANC) is defined as follows:
\begin{equation}
    R(v_1,v_2,\cdots,v_N)=\frac{1}{N}\sum_{k=1}^{N}\frac{\sigma(\mathcal{G}\backslash\{v_1,v_2,\cdots,v_k\})}{\sigma(\mathcal{G})}\text{,}
\end{equation}
where $N$ is the total number of the nodes,and $\sigma$ is the widely used pairwise connectivity.The $\sigma:\{\mathcal{G}\}\to\{R^+\}$ is defined as:
\begin{equation}
    \sigma(G)=\sum_{C_i\in G}\frac{|C_i|(|C_i|-1)}{2}\text{,}
\end{equation}
where $C_i$ is the $i$th connected component in network G,and $|C_i|$ means the number of nodes in the subgraph $C_i$. Our goal is to find a removal list that can minimize the ANC of a given network after removing selected nodes.
\subsubsection{Baselines}
We compare our proposed method with the following competitive baselines:

\begin{itemize}[leftmargin=*]
    \item \textbf{DC (Degree Centrality)}: A heuristic method that iteratively removes nodes with the highest degrees.
    \item \textbf{GNDR} \cite{GND:ren2019generalized}: Employs a spectral approach to iteratively partition the network into two components.
    \item \textbf{MinSum} \cite{minsum:braunstein2016network}: Initially creates an acyclic network by removing nodes that form loops, then employs greedy tree-breaking algorithms to decompose the resultant forest into smaller, disconnected components.
    \item \textbf{CoreHD} \cite{CoreHD:zdeborova2016fast}: Utilizes a \textit{k}-core decomposition strategy, where the \textit{k}-core of a network denotes the maximal subgraph in which every node has a degree of at least k. CoreHD progressively removes nodes with the highest degrees within its \textit{k}-core to select key nodes.
    \item \textbf{WN (Weak Neighbors)} \cite{WeakNeighbors:schmidt2019minimal}: This algorithm improves upon CoreHD by selecting, for deletion, the neighbor node with the minimum degree when multiple nodes have the maximum degree in a \textit{k}-core. Conversely, CoreHD randomly selects one of the nodes with the maximum degree within a \textit{k}-core for deletion.
    \item \textbf{VE} \cite{VE:huang2024identifying}: Incorporating concepts from quantum mechanics, VE utilizes an entanglement-based metric to quantify the perturbations that individual vertices induce in spectral entropy, and it identifies key nodes based on these measurements.
    \item \textbf{GDM} \cite{GDM:grassia2021machine}: This machine learning-based approach uses graph convolutional networks to understand the topological structure of networks and to evaluate the likelihood of node attacks. It is trained on small-scale virtual networks, showing favorable transferability to large-scale and real-world networks.
    \item \textbf{NIRM} \cite{NIRM:zhang2022dismantling}: This machine learning-based method conducts a comprehensive assessment of both global and local node features in the network, effectively identifying critical nodes.
\end{itemize}

\begin{table}[h]
    \vspace{-1em} 
    \centering
    \caption{Statistics of networks}
    \begin{tabular}{c|cccc}
    \toprule
         Network&Jazz&Twitter&Network Science&Synthetic Network\\
    \midrule
        $|V|$&$198$&$475$&$1,565$&$1,000$\\
        $|E|$&$2,742$&$13,289$&$13,532$&$2,991$\\
    \bottomrule
    \end{tabular}
    \label{tab:description of networks}
\end{table}
\subsubsection{Datasets}
We selected the following datasets to evaluate our method and baselines, the statistics of which are detailed in Table \ref{tab:description of networks}.

\begin{itemize}[leftmargin=*]
    \item Jazz \cite{jazz:gleiser2003community}: Represents a collaboration network among Jazz musicians, where each node corresponds to a musician, and an edge signifies that two musicians have collaborated in a band.
    \item Twitter \cite{twitter:fink2023twitter}: This dataset quantifies the pairwise probability of influence within a Congressional Twitter network.
    \item Network-Science \cite{networkscience:newman2006finding}: Constitutes a co-authorship network within the field of network science.
    \item Synthetic Network: A Barabási-Albert $(m=3)$ network, randomly generated with 1000 nodes.
\end{itemize}

\subsection{Overall Performance Comparison (RQ1)} 

\begin{table*}[t]
\caption{Overall performance of our method. The ANC presented here represents  ANC score of networks post the removal of 20\% of nodes selected by each method and a lower ANC indicates greater efficacy.}
\label{tab:overall performance}
\scriptsize
\begin{tabular}{c|cc|cc|cc|cc|c}
\toprule
\multirow{2}{*}{\diagbox{Methods}{Datasets}} & \multicolumn{2}{c|}{Jazz} & \multicolumn{2}{c|}{Network Science} & \multicolumn{2}{c|}{Twitter} & \multicolumn{2}{c|}{Synthetic Network}&\multirow{2}{*}{Average Rank} \\
&ANC&rank&ANC&rank&ANC&rank&ANC&rank \\
\midrule
LLMs+EA & \colorbox{yellow!15}{0.65454} &\colorbox{yellow!15}{1}& \textbf{0.05326}&\textbf{2} & \textbf{0.80762}& \textbf{2}& \textbf{0.65495}&\textbf{2}&\colorbox{yellow!15}{1.75}\\
\midrule
DC& 0.79066&8 & 0.11842&9 & 0.81125&4 & 0.67685&5 &6.5\\
\midrule
GNDR & \textbf{0.69697} &\textbf{2}& 0.05330&3 & 0.80799&3 & 0.70592&7 &3.75\\
MinSum & 0.79414&9 & \colorbox{yellow!15}{0.05291}&\colorbox{yellow!15}{1} & \colorbox{yellow!15}{0.80535}&\colorbox{yellow!15}{1} & 0.65613&3&3.5 \\
CoreHD & 0.78833&5 & 0.09932&7 & 0.81125&4 & 0.73633&8&6 \\
WN & 0.78833 &5& 0.10325&8 & 0.81125&4 & 0.73795&9& 6.5\\
VE & 0.74837&3 & 0.08583&6 & 0.81125&4 & 0.68259&6&4.75 \\
\midrule
GDM & 0.78417&4 & 0.05908&4 & 0.81125&4 & \colorbox{yellow!15}{0.65384}&\colorbox{yellow!15}{1} &\textbf{3.25}\\
NIRM & 0.78855&7 & 0.06781&5 & 0.81125&4 & 0.67446&4&5 \\
\bottomrule
\multicolumn{10}{l}{Data highlighted with \colorbox{yellow!15}{yellow} means its method performs the best on the network, while the \textbf{bolded} data's method is the second best.}
\end{tabular}
\vspace{-1em} 

\end{table*}

First, we implement our proposed model and baseline methods on each network dataset, subsequently selecting 20\% of the total nodes from each network. Following this, we exclude these nodes from the network and compute the ANC for each case. Table \ref{tab:overall performance} presents a comparison of the overall performance. Based on these results, we make the following observations:

\begin{itemize}[leftmargin=*]
\item \textbf{The performance of baselines varies across networks.} 

The effectiveness of baseline methods differs significantly among networks, where a lower ANC value reflects a more substantial impact on network connectivity. Specifically, GNDR outperforms others in the Jazz network, followed by the learning-based method VE, with the remaining approaches closely trailing. Notably, MinSum shows the least effectiveness here. Conversely, in the Network Science dataset, the generally low ANC values suggest a pronounced effect on network connectivity by all baseline methods, with MinSum leading in performance and the heuristic DC lagging behind. In the Twitter network, MinSum continues to excel, although the ANC values for all methods are tightly grouped, indicating less differentiation in the identification of critical nodes. Remarkably, six baselines yield identical ANC values in this context, underscoring the distinct nature of critical nodes within the Twitter network. For the synthetic network, GDM emerges as the most effective, whereas Weak-Neighbors falls short.

\item \textbf{Superior Overall Performance of Our Model.}

Our method distinguishes itself by surpassing all competitors in the Jazz dataset by a notable margin and maintaining robust performance in the Network Science, Twitter, and synthetic networks, consistently ranking as a close second to the best-performing methods. Interestingly, MinSum, despite its top-tier results in the Network Science and Twitter datasets, ranks at the bottom in the Jazz dataset, highlighting its inconsistent efficacy across different networks. The uniform excellence of our model across diverse networks suggests its capability to more accurately score nodes, thereby more effectively identifying critical nodes within any given network.
\end{itemize}
\subsection{Function Analysis (RQ2)}

\begin{figure*}[t]
  \centering
  \includegraphics[width=1\textwidth]{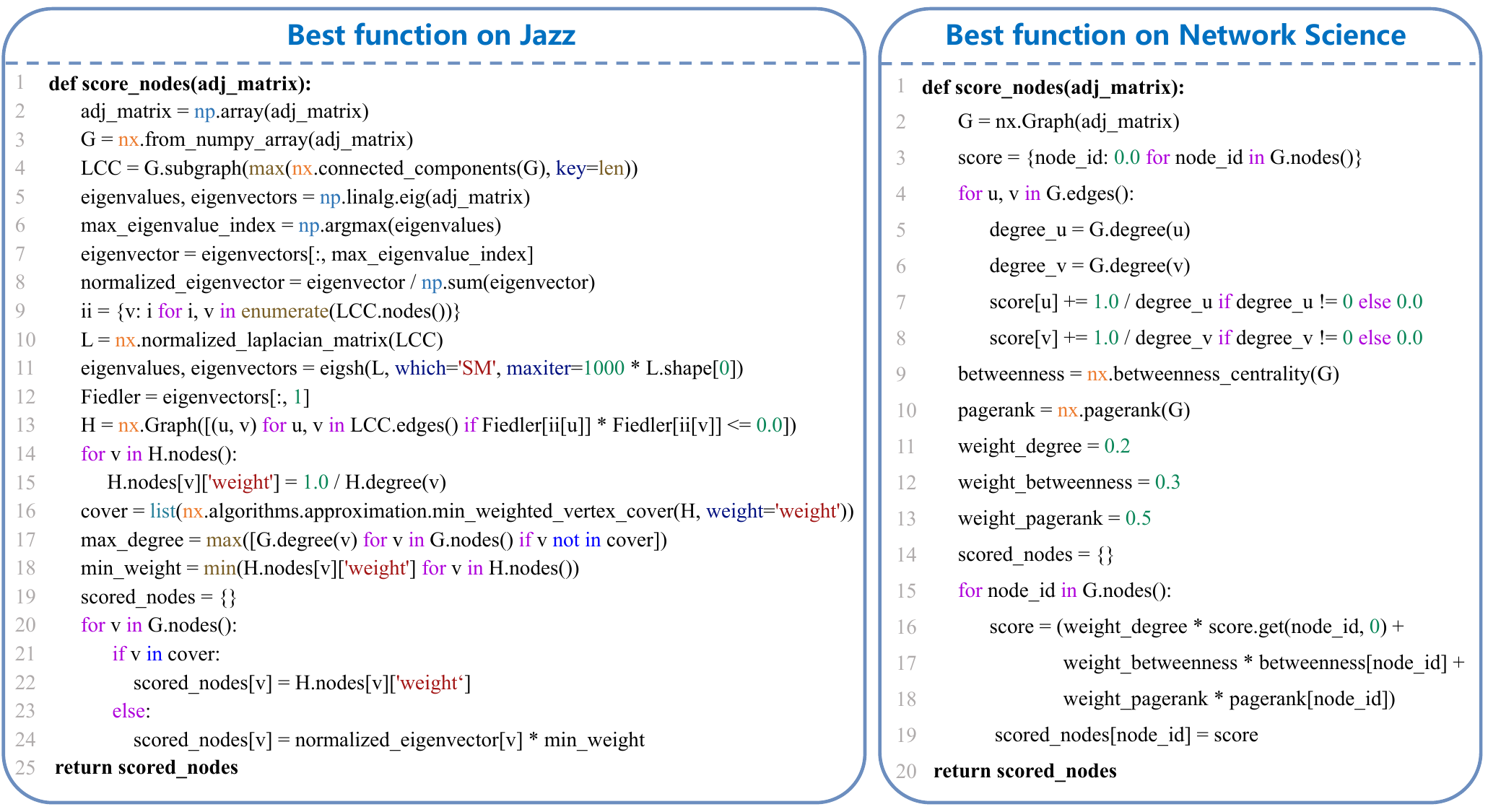}
  \caption{The best functions found on Jazz and Network-Science.} 
  \label{fig::best functions}
  \vspace{-1em} 
\end{figure*}

We present the optimal node scoring functions derived from the Jazz and Network-Science datasets in Figure \ref{fig::best functions}. For Jazz, the premier scoring function assesses nodes by integrating local features with centrality measures. It begins by computing the network's largest connected component to derive a Laplacian matrix (lines 4, 9–10). Utilizing this matrix, the function proceeds to compute the Fiedler vector, employing it to establish a new graph (lines 11–13). This new graph identifies a minimum weighted cover to assign scores to its nodes (lines 13–22). Nodes not included in this new graph receive scores based on normalized eigenvector centrality (lines 5–8, 24).

Conversely, the superior node scoring function from the Network-Science dataset determines node scores by calculating degree-related metrics (lines 4–8), betweenness centrality (line 9), and PageRank values (line 10) independently. These metrics are then aggregated using weighted averages to calculate a comprehensive node score (lines 11–19), effectively merging various centrality measures for node evaluation.

This analysis reveals that the scoring functions our method uncovers differ across networks, all originating from manually designed initial functions and undergoing diverse modifications to logically score nodes. Additional discovered functions are accessible here: \url{https://anonymous.4open.science/r/LLM4CN-6520}.




\begin{figure*}[b]
\vspace{-0.3in}
  \centering
    \subfloat[Population Size Heatmap]{\includegraphics[width=0.33\textwidth]{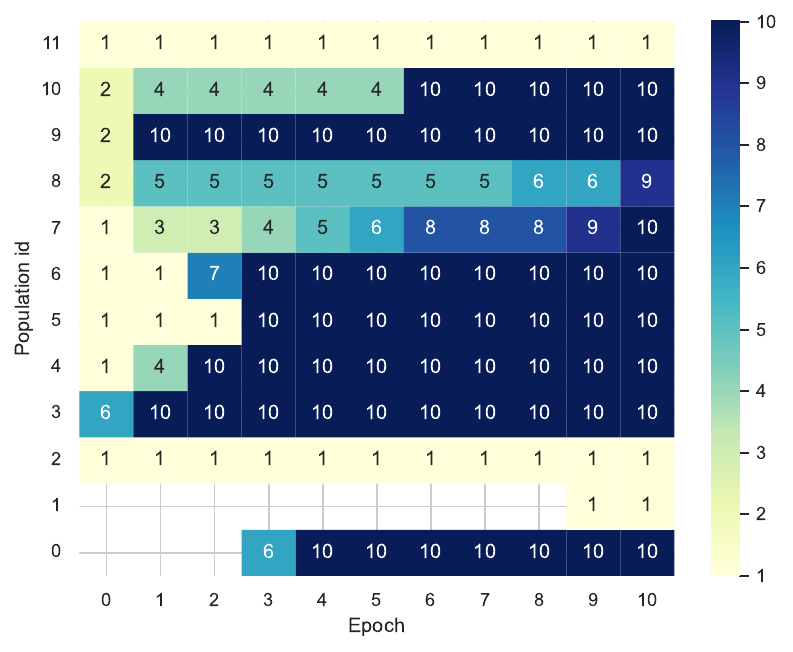}}
    \subfloat[Average Fitness Score Heatmap]{\includegraphics[width=0.33\textwidth]{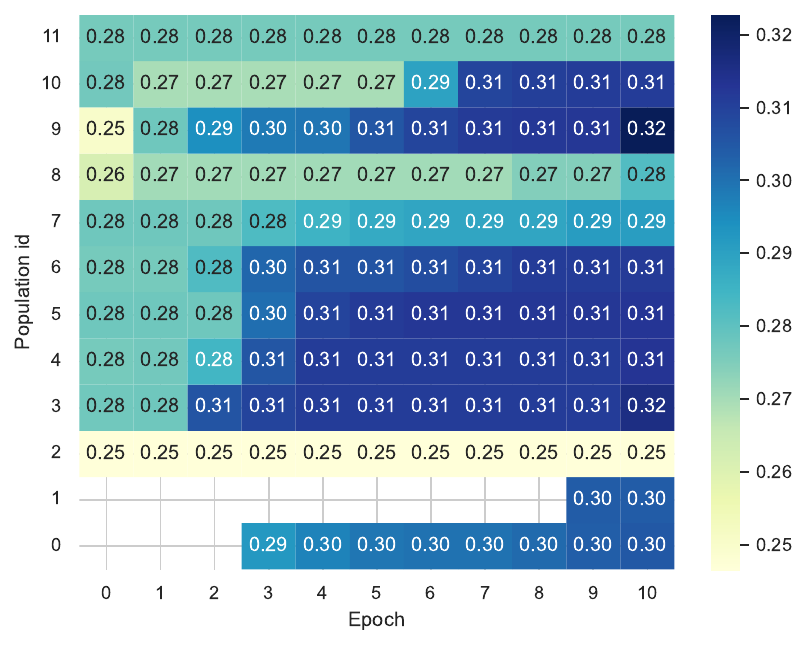}}
    \subfloat[Highest Fitness Score Heatmap]{\includegraphics[width=0.33\textwidth]{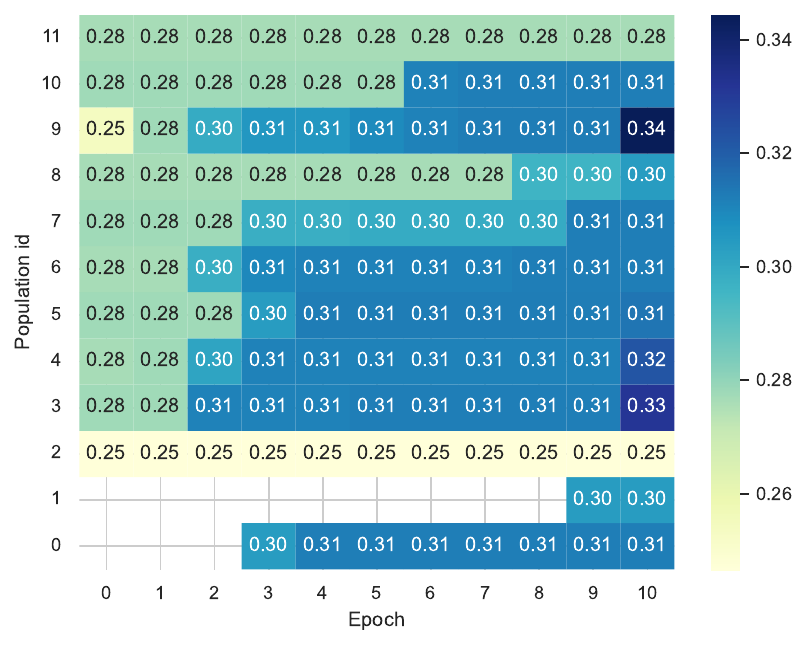}}
  \setlength{\abovecaptionskip}{0.2cm}
  \setlength{\belowcaptionskip}{0.2cm}
  \caption{Evolution of funciton populations.} 
  \label{fig::evolution}
\end{figure*}

\subsection{Case Study (RQ3)}

Figure \ref{fig::evolution} illustrates the statistics of function populations during the training process on the synthetic network, aiding us in understanding the evolutionary process of the function populations. Figure \textbf{(a)}, \textbf{(b)} and \textbf{(c)} respectively represent the evolution of the population size, the average fitness score, and the highest fitness score of each function.

First, in terms of population size, as evolution progresses, the majority of populations incrementally increase, reaching their respective capacities and initiating updates via elimination. However, for Population 2, the function count consistently remains at one, indicating the absence of similar offspring function generation. Given that Population 2's score is the lowest among all initial functions, it logically follows that its \textit{inferior genes} are not selected.  Furthermore, two new populations emerged during evolution, illustrating our population management design's capability to preserve the diversity of function populations.

Moreover, the analysis of change curves for average and highest fitness scores across populations reveals continuous enhancement in the overall function quality.
Simultaneously, no instances of multiple populations maintaining the same average fitness throughout evolution were observed, indicating our design's effectiveness in circumventing the pitfall of convergent evolution. Around the 20th and 100th epochs, there are noticeable increases in the fitness scores of the function populations, representing the circulation of \textit{superior genes} within the populations and enhancing overall performance.

In conclusion, our proposed model ensures that the function populations can evolve continuously while preserving diversity, thereby facilitating the generation of superior new functions.

\subsection{Ablation Study (RQ4)}

We conduct ablation studies to evaluate the effectiveness of a node scoring function in identifying critical nodes within a network. Our model is built upon three key components, and through ablation, we aim to assess the impact of each component as described below.
\begin{wrapfigure}{r}{0.4\textwidth}
  \begin{center}
    \includegraphics[width=0.4\textwidth]{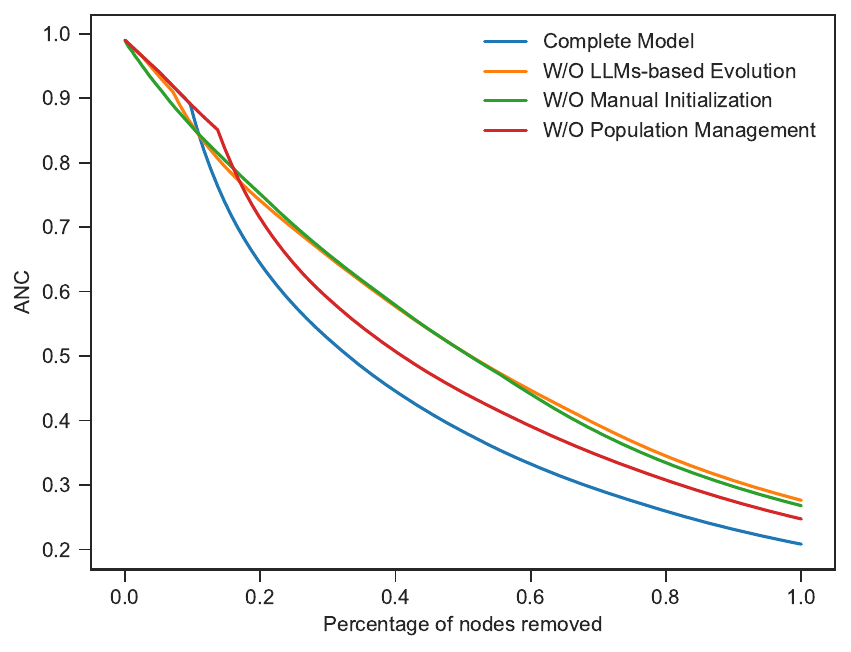}
  \end{center}
\vspace{-0.1in}
\caption{Comparison of ablation study.}
    \label{fig::ablation study}
\vspace{-1em} 
\end{wrapfigure}

\begin{itemize}[leftmargin=*]
    \item \textbf{W/O Manual Initialization:} To gauge the significance of Manual Initialization, we adopt an alternate strategy. Here, we define the task of identifying critical nodes to the LLM and instruct it to generate an equivalent number of node scoring functions in a predetermined format, thereby bypassing manual initialization.

    \item \textbf{W/O Population Management:} To explore the necessity of Population Management, we implement an alternative scheme where all functions are collected without any categorization. During the selection phase, parent functions are randomly selected from the entire pool of individuals, eliminating structured population management.
    \item \textbf{W/O LLMs-based Evolution:} Assessing the indispensability of LLMs-based Evolution, we introduce an alternative method. Initially, all functions are treated as parent functions. This is followed by a single epoch of crossover and mutation to produce a multitude of offspring functions, from which the most suitable are chosen, simulating a condensed version of the evolutionary process.
\end{itemize}
The efficacy of the node scoring functions discovered via each ablation strategy versus the comprehensive method was analyzed. Findings, depicted in Figure \ref{fig::ablation study}, affirm that every component of our proposed model significantly enhances its overall performance.

\section{Related Work}\label{sec::relatedwork}

\subsection {Synergy between LLMs and EAs}
Recent research underscores the effectiveness of leveraging Large Language Models (LLMs) for optimization tasks via prompt engineering \cite{yang2023large, guo2023towards}. This method proves particularly advantageous for Evolutionary Algorithms (EAs), attributed to the advanced text comprehension and generation capabilities of LLMs \cite{chao2024match, wu2024evolutionary}. For example, recent investigations into discrete prompt optimization have shown how LLMs can mimic evolutionary operators, generating innovative candidate prompts through crossover and mutation based on fixed prompts \cite{guo2023connecting, li2023spell}. Furthermore, studies have explored the use of LLMs in tackling combinatorial optimization problems (COPs) by developing evolved heuristic solutions \cite{pluhacek2023leveraging, ye2024reevo}. EvoPrompting \cite{chen2023evoprompting} utilizes LLMs for Neural Architecture Search (NAS), performing crossover and mutation at the code level to discover new architectures. Likewise, FunSearch \cite{romera2023mathematical} implements an LLM-powered evolutionary method to determine solution functions for mathematical challenges. In addition, Eureka \cite{ma2023eureka} focuses on the design of reinforcement learning reward functions using LLM-based evolutionary algorithms, while Fuzz4All \cite{fuzz4all} applies LLMs in fuzz testing to generate and mutate inputs.

Distinct from previous studies, our research emphasizes the use of LLMs that perform mutation and crossover processes on node scoring functions, aiming to evolutionarily determine the most effective scoring functions.


\vspace{0.1cm}
\subsection{Critical Nodes Detection}

Identifying critical nodes is of significant importance across various practical domains. Numerous approaches have been proposed to tackle this issue.
Basic methods such as Degree Centrality, entail removing nodes with the highest degrees.
Algorithms like CoreHD \cite{CoreHD:zdeborova2016fast} and WeakNeighbors \cite{WeakNeighbors:schmidt2019minimal} utilize \textit{k}-core decomposition, in which the \textit{k}-core of a network is defined as the maximal subgraph with every node having a degree of at least k. These methods progressively eliminate nodes with the highest degrees within their respective \textit{k}-cores to identify key nodes. On the other hand, Min-Sum \cite{minsum:braunstein2016network} focuses on creating an acyclic network by removing nodes that form loops, followed by utilizing greedy tree-breaking algorithms to decompose the remaining forest into smaller disconnected components. Additionally, GND \cite{GND:ren2019generalized} adopts a spectral approach to iteratively partition the network into two components. Learning-based methods like NIRM \cite{NIRM:zhang2022dismantling} and GDM \cite{GDM:grassia2021machine} utilize neural networks to assign scores to network nodes, determining their importance based on these scores.

Unlike previous studies, our research is not limited to identifying specific algorithms. Rather, it leverages LLM-empowered Evolutionary Algorithms (EAs) to investigate a broad spectrum of potential algorithms, yielding favorable outcomes across diverse networks.

\vspace{0.1cm}
\subsection{LLMs for Code Generation}

Large Language Models (LLMs) exhibit considerable promise in automating code generation, capitalizing on their capacity for training on vast repositories of publicly available source code. This training enables them to acquire a comprehensive understanding of programming concepts and patterns, which in turn significantly enhances their capabilities in code generation \cite{chen2021evaluating, li2022competition}. These capabilities are progressively applied to tackle a multitude of code-related challenges. Notably, LLMs are employed in tasks such as code debugging \cite{chen2023teaching, liventsev2023fully}, enhancing code performance \cite{madaan2023learning}, resolving algorithmic competition problems \cite{li2022competition, pmlr-v202-guo23j}, and conducting code deobfuscation \cite{armengol2023slade, xu2023lmpa} and translation \cite{pan2023understanding, tang2023explain}. Additionally, the capacity to produce task-specific code highlights the versatility of LLMs; for instance, their deployment in robotics control \cite{liang2023code, wang2023gensim}, domain-specific data curation \cite{chen2023seed}, and solving general tasks \cite{zhang2022planning, yang2023intercode}, illustrates their extensive utility across specialized and general domains.

In this paper, we explore in depth the potential of code generation in efficiently identifying optimal scoring functions through the use of LLM-based evolutionary algorithms (EAs).

\section{Conclusion and Future Work}\label{sec::conclusion}

In this study, we employ an evolutionary algorithm enhanced by large language models for generating node scoring functions, to identify critical nodes within networks. Initially, the evolution begins with a set of functions crafted based on network topology and existing algorithmic approaches. Subsequently, the model assesses the fitness scores of the given functions within the network and categorizes populations according to the embedding representations of these functions. Following this, selections both between and within populations are performed to select parent algorithms for further evolution. Finally, the LLMs are utilized to execute crossover and mutation operations, producing offspring individuals. Following executable checks, newly generated functions are incorporated into the population, and this iterative process persists until the evolution terminates.

Experimental results across various network datasets reveal that our method effectively balances adaptability and utility in comparison to other state-of-the-art algorithms. It is capable of generating a variety of both reasonable and efficient node scoring functions.

In future work, we aim to delve deeper into distributed designs for the evaluation, management, and evolution processes, enabling parallel execution to improve the model’s evolutionary speed and performance. Furthermore, we plan to diversify the types of initial algorithms and prompts, aiming to create more compelling, high-performance node scoring functions through large language models, focusing on the identification of critical nodes within networks.

\bibliographystyle{plain}
\bibliography{references}

\end{document}